\documentclass[%
 reprint,
 amsmath,
 amssymb,
 aps,
 pra,
twocolumn,
]{revtex4-1}

\usepackage{graphicx}
\usepackage{dcolumn}
\usepackage{bm}
\usepackage{gensymb}
\usepackage{physics}
\usepackage{color}
\usepackage{tikz}
\usepackage{soul}
\usepackage{upgreek}
\usepackage{makecell}
\usepackage{multirow}
\usepackage{hyperref}

\hypersetup{
    colorlinks=true,       
    linkcolor=blue,          
    citecolor=blue,        
    filecolor=blue,      
    urlcolor=blue           
}

\usepackage{txfonts}

\newcommand{\term}[3]{\phantom{ }^{#1}#2_{#3}}

\def\ms{\mbox{ ms}}
\def\us{\mbox{ $\mu$s}}
\def\um{\mbox{ $\mu$m}}
\def\nm{\mbox{ nm}}
\def\mm{\mbox{ mm}}

\def\Hz{\mbox{ Hz}}
\def\kHz{\mbox{ kHz}}
\def\MHz{\mbox{ MHz}}
\def\us{\mbox{ $\mu$s}}
\def\mW{\mbox{ mW}}

\newcommand\T{\rule{0pt}{2.6ex}}       

\begin{document}

\preprint{APS/123-QED}

\title{A 3-photon process for producing a degenerate gas of metastable alkaline-earth atoms}

\author{D.S. Barker}
\email{dbarker2@umd.edu}
\author{N.C. Pisenti}
\author{B.J. Reschovsky}%
\author{G.K. Campbell}
\affiliation{%
 Joint Quantum Institute, University of Maryland and National Institute of Standards and Technology, College Park, MD 20742
}%

\date{\today}

\begin{abstract}

We present a method for creating a quantum degenerate gas of metastable alkaline-earth atoms.
This has yet to be achieved due to inelastic collisions that limit evaporative cooling in the metastable states.
Quantum degenerate samples prepared in the $\term{1}{S}{0}$ ground state can be rapidly transferred to either the $\term{3}{P}{2}$ or $\term{3}{P}{0}$ state via a coherent 3-photon process.
Numerical integration of the density matrix evolution for the fine structure of bosonic alkaline-earth atoms shows that transfer efficiencies of $\simeq90\%$ can be achieved with experimentally feasible laser parameters in both Sr and Yb.
Importantly, the 3-photon process can be set up such that it imparts no net momentum to the degenerate gas during the excitation, which will allow for studies of metastable samples outside the Lamb-Dicke regime.
We discuss several experimental challenges to successfully realizing our scheme, including the minimization of differential AC Stark shifts between the four states connected by the 3-photon transition.

\end{abstract}

\maketitle


\section{\label{intro}Introduction}

Alkaline-earth-like (AE) atoms have attracted experimental and theoretical interest due to their narrow optical resonances and non-magnetic ground state.
Recent experiments have exploited these properties to study atom interferometry~\cite{Jamison2014a,Tarallo2014,Sorrentino2009}, atom clocks~\cite{Hinkley2013,Bloom2014}, superradiant lasers~\cite{Norcia2016}, quantum simulation~\cite{Hofrichter2015,Yamamoto2016,Nakajima2016}, and molecular physics~\cite{Stellmer2012a,Reinaudi2012,Mcguyer2013,McDonald2015}.
An outstanding experimental challenge is the realization of quantum degenerate samples of AE atoms in the metastable $\term{3}{P}{2}$ and $\term{3}{P}{0}$ states.
These samples would be useful in a wide variety of applications.
For example, the $\term{3}{P}{2}$ state has a permanent electric quadrupole moment, and $\term{3}{P}{2}$ degenerate gases are a potential platform for quantum simulation~\cite{Bhongale2013} or for studies of anisotropic collisions~\cite{Derevianko2003,Lahrz2014}.
Degenerate samples of $\term{3}{P}{0}$ atoms could help to advance atomic structure calculations~\cite{Herold2012,Holmgren2012,Safronova2015}, increase the accuracy of atomic clocks~\cite{Nicholson2015}, or generate highly entangled states~\cite{Foss-Feig2012}.
Simultaneous coherent manipulation of atoms in both $\term{3}{P}{2}$ and $\term{3}{P}{0}$ is required for several proposed quantum computing schemes~\cite{Gorshkov2009,Daley2008a}.

Inelastic collisional losses, which are on the order of $10^{-10}-10^{-11}~\text{cm}^{3}/\text{s}$, prevent direct evaporation of metastable AE atoms to quantum degeneracy~\cite{Traverso2009, Lisdat2009, Yamaguchi2008,Uetake2012,Halder2013}.
Previous experiments used either incoherent excitation~\cite{Khramov2014a,Yamaguchi2008,Traverso2009,Halder2013} or coherent excitation of a doubly-forbidden transition~\cite{Zhang2014,Uetake2012,Hara2014} to transfer pre-cooled AE atoms to a metastable state.
These single-photon techniques necessarily impart momentum to the sample during excitation, which limits their application to either thermal atoms or to the Lamb-Dicke regime.
Additionally, addressing a doubly-forbidden transition is challenging due to the stringent requirements on the excitation laser's linewidth and the need for accurate spectroscopy of the $\term{1}{S}{0}\rightarrow\term{3}{P}{0\,(2)}$ transitions, which has only been performed on a few isotopes of AE atoms~\cite{Hoyt2005,Campbell2008,Yamaguchi2010,Morzynski2015,Dareau2015}.

\begin{figure}
\includegraphics[width=\linewidth]{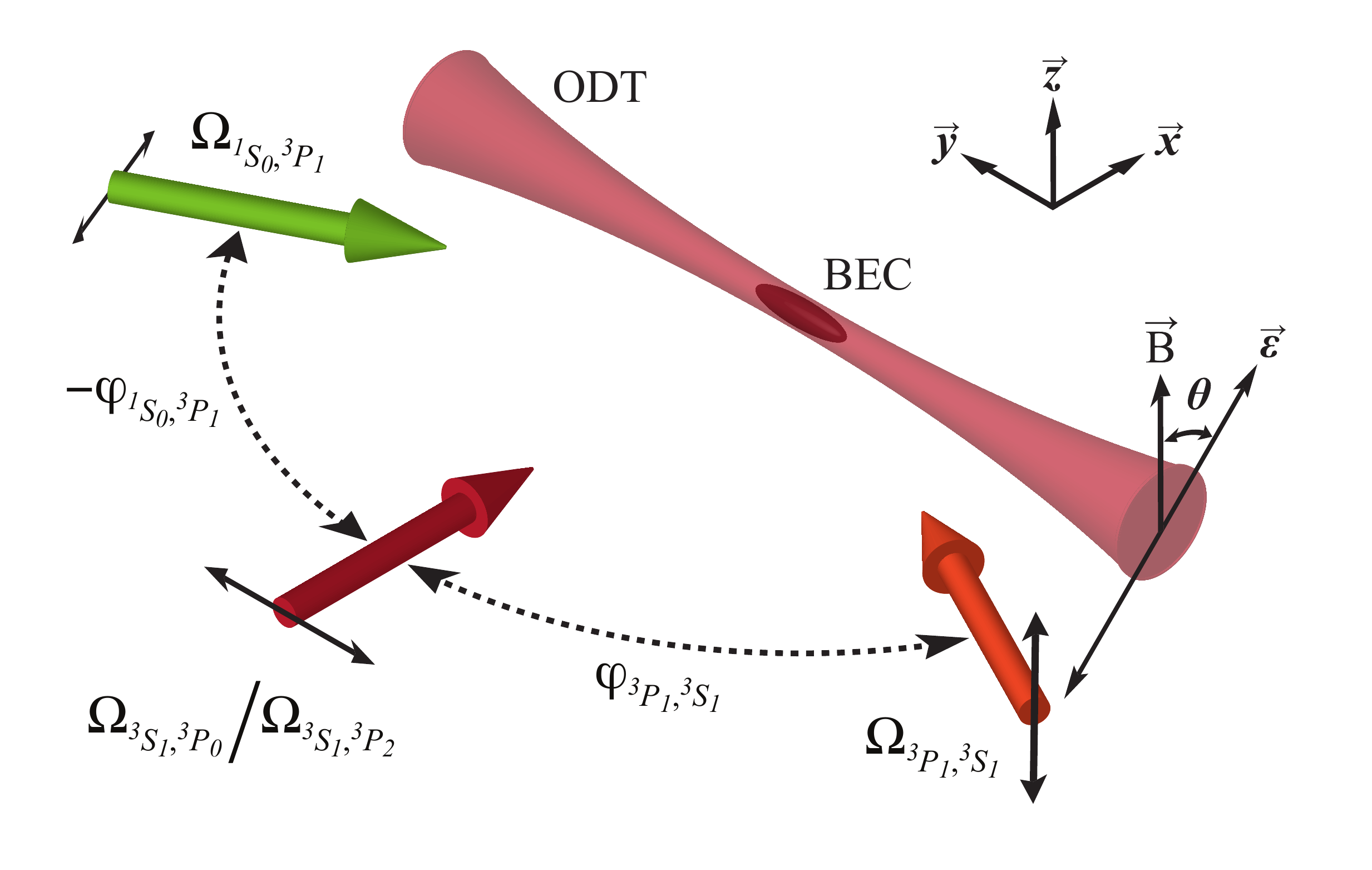}
\caption{\label{scheme}(Color Online) Laser configuration for the 3-photon excitation scheme. A Bose-Einstein Condensate (BEC) sits in an Optical Dipole Trap (ODT) in a region with a uniform magnetic field, $\vec{B}=B\,\hat{z}$. Three lasers cross at the BEC and drive the transition to the metastable state. These lasers address the $^{1}S_{0}\rightarrow\, ^{3}P_{1}$, $^{3}P_{1}\rightarrow\, ^{3}S_{1}$, and $^{3}S_{1}\rightarrow\, ^{3}P_{0\,(2)}$ transitions with Rabi frequencies of $\Omega_{^{1}S_{0},^{3}P_{1}}$, $\Omega_{^{3}P_{1},^{3}S_{1}}$, and $\Omega_{^{3}S_{1},^{3}P_{0\,(2)}}$, respectively. The angles of incidence, $\upvarphi_{^{3}P_{1},^{3}S_{1}}$ and $\upvarphi_{^{1}S_{0},^{3}P_{1}}$, between the excitation laser beams and the $x$-axis can be chosen so as to eliminate the net momentum transfer to the condensate during the excitation process. The double headed black arrows indicate the polarization of the optical fields, which decomposes to either $|\sigma=1\rangle\,{+}\,|\sigma={\mbox{\small{-}}}1\rangle$ or $|\sigma=0\rangle$ in the circular basis, depending on whether the laser is polarized parallel or perpendicular to the magnetic field. The angle between the ODT polarization vector and the magnetic field, $\theta$, can be tuned to control dephasing due to differential AC Stark shifts between the four atomic states (see Section~\ref{exp}).}
\end{figure}

We propose a 3-photon excitation scheme for the creation of degenerate gases in metastable states, which transfers the atoms through the $\term{1}{S}{0}\rightarrow\term{3}{P}{1}\rightarrow\term{3}{S}{1}\rightarrow\term{3}{P}{0\,(2)}$ path.
During the 3-photon process each atom absorbs two photons and emits one, so an appropriate laser arrangement can eliminate the net momentum transfer to the atomic sample (see Figure~\ref{scheme}).
All the transitions addressed in this scheme are much broader than the doubly-forbidden transitions in AE atoms (the smallest single-photon linewidth is $\simeq370\Hz$ for Ca, $\simeq7.5\kHz$ for Sr, and $\simeq180\kHz$ for Yb), which substantially relaxes the laser linewidth requirements.
For the 3-photon process to be coherent, the three lasers must be phase-locked.
Although the necessary wavelengths potentially span hundreds of nanometers (see Table~\ref{lambda}), the lasers can be stabilized to each other with a cavity transfer lock, an electromagnetically induced transparency (EIT) lock, or a beatnote lock to an optical frequency comb~\cite{Bell2007,Abel2009,Liekhus-Schmaltz2012,Inaba2013,Scharnhorst2015,Bridge2016}. 

Here, we investigate the feasibility of the 3-photon scheme by numerically integrating the Optical Bloch Equations (OBEs) for the 13-level system of bosonic AE atoms (see Figure~\ref{levels}).
The simulations use linewidths and wavelengths for strontium and ytterbium, but the generic results should apply to calcium as well.
We present the numerical results and estimate the effective 3-photon linewidth in Section~\ref{thry}.
Section~\ref{exp} describes possible solutions to several experimental challenges, including ways to extend the lifetime of the metastable sample and mitigate inhomogeneous broadening.
We summarize our results and discuss future outlook in Section~\ref{con}.

\section{\label{thry}3-photon dynamics}

\begin{figure}
\includegraphics[width=\linewidth]{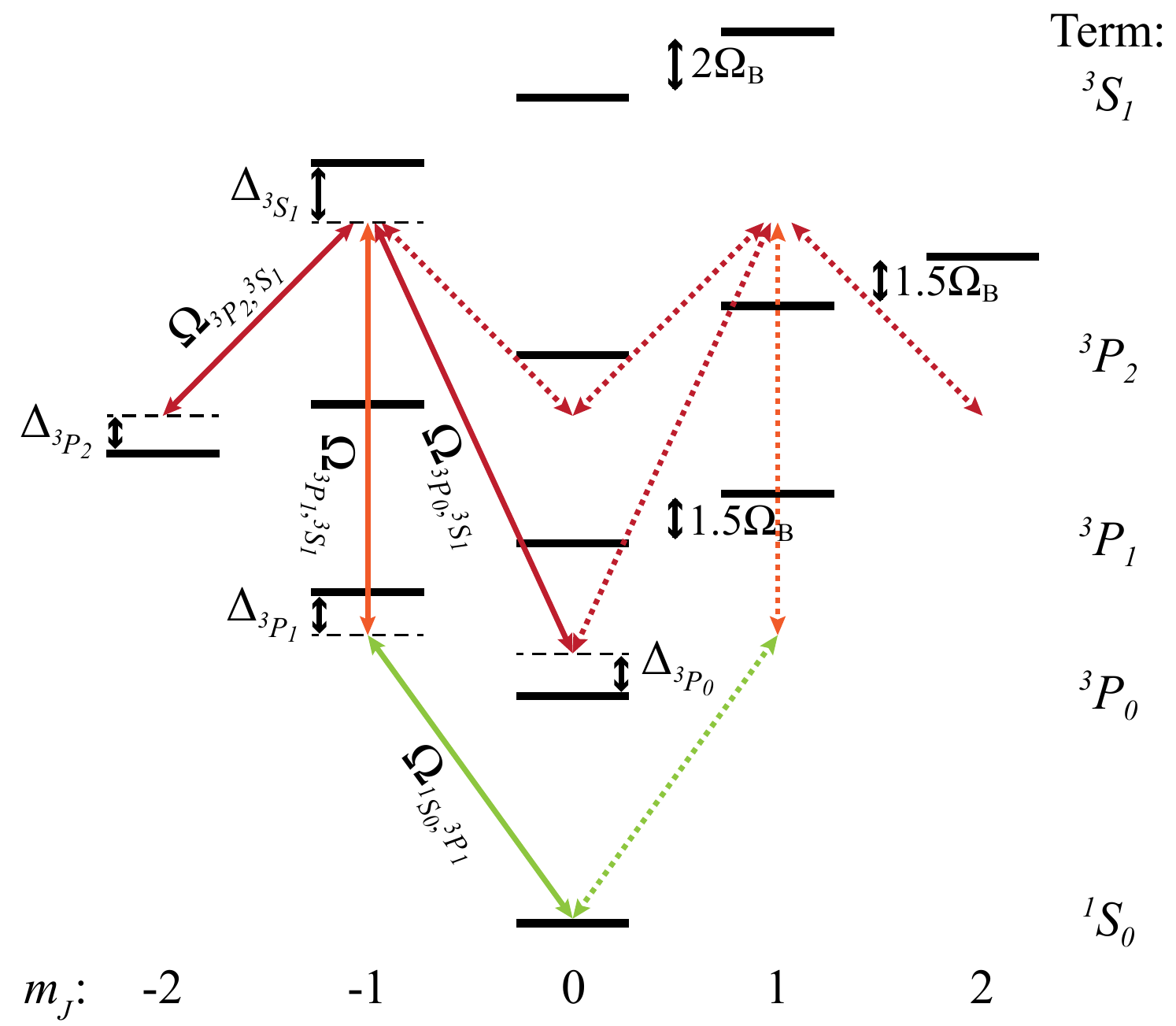}
\caption{\label{levels}(Color Online) The Zeeman level structure of bosonic AE atoms relevant for the 3-photon process. The colored double-headed arrows indicate dipole-allowed transitions for each optical field after projection of its polarization onto the quantization axis. These arrows are solid for the transitions closest to resonance and dotted for transitions that are far off resonance. The polarization projection depends on the angle, $\alpha_{\ell,\ell'}$, between the laser polarization vector and the magnetic field, $\vec{B} = \Omega_{B}\hat{z}/\mu_{B}$ (see Figure~\ref{scheme}), where $\ell$ ($\ell'$) is the term symbol of the lower (upper) state of the transition and we have taken $\hbar=1$. The excitation laser strength is $E_{\ell,\ell'}=\Omega_{\ell,\ell'}/\langle\ell'\|\,\vec{d}\,\|\ell\rangle$ with $\langle\ell'\|\,\vec{d}\,\|\ell\rangle$ the reduced dipole matrix element. The detuning, $\Delta_{\ell}$, is referenced to the lowest energy Zeeman state of $|\ell\rangle$. Here, we have depicted a 3-photon transition to either $|^{3}P_{2},-2\rangle$ or $|^{3}P_{0},0\rangle$, but $|^{3}P_{2},2\rangle$ can be reached by reversing the magnetic field or detuning each laser to the blue of the highest energy Zeeman state that it addresses. RF transitions from $|^{3}P_{2},\pm2\rangle$ can prepare other Zeeman states in $^{3}P_{2}$.}
\end{figure}

Figure~\ref{levels} shows the 13 relevant Zeeman sublevels for bosonic AE atoms and the coupling lasers needed for the 3-photon transition.
We label the states $|\ell,m_{J}\rangle$, where $\ell=\term{2S+1}{L}{J}$ is the term symbol for the state and $m_{J}$ the projection of the total electronic angular momentum, $J$, onto the $z$-axis.
The optical field coupling level $|\ell\rangle$ to $|\ell'\rangle$ is given by its electric field magnitude, $E_{\ell,\ell'}=\Omega_{\ell,\ell'}/\langle\ell'\|\,\vec{d}\,\|\ell\rangle$, where $\Omega_{\ell,\ell'}$ is the single-photon Rabi frequency, $\langle\ell'\|\,\vec{d}\,\|\ell\rangle$ is the reduced dipole matrix element, and we have taken $\hbar=1$.
The 1-photon laser detuning is $\Delta_{\term{3}{P}{1}}$, the 2-photon detuning is $\Delta_{\term{3}{S}{1}}$, and the 3-photon detuning is either $\Delta_{\term{3}{P}{2}}$ or $\Delta_{\term{3}{P}{0}}$ depending on the desired final state.
All the detunings, $\Delta_{\ell}$, are referenced to the lowest energy Zeeman state of level $|\ell\rangle$.
A magnetic field, $\vec{B} = \Omega_{B}\hat{z}/\mu_{B}$ with $\mu_{B}$ the Bohr magneton, breaks the degeneracy of the Zeeman states.
This splitting allows individual addressing of $\term{3}{P}{0}$ or the $m_{J}={\pm}2$ states of $\term{3}{P}{2}$.
For our calculation, we selected $|\term{3}{P}{2},\pm2\rangle$ as the target states in $\term{3}{P}{2}$ based on their large quadrupole moment~\cite{Bhongale2013} and favorable Clebsch-Gordan overlap with $|\term{3}{S}{1},\pm1\rangle$.
Other $\term{3}{P}{2}$ Zeeman levels can be prepared using RF transitions, as has been demonstrated in~\cite{Khramov2014a}.
The laser polarizations, Zeeman splittings, and laser detunings must be carefully chosen to suppress dipole-allowed transitions to undesired states (see Figure~\ref{levels}).
For example, the laser driving the $|\term{3}{P}{1}\rangle\rightarrow|\term{3}{S}{1}\rangle$ transition must be $\pi$ polarized to prevent unwanted accumulation of atoms in $|\term{3}{P}{1},\pm1\rangle$.

In the rotating wave approximation, the Hamiltonian for the system shown in Figure~\ref{levels} is~\cite{Steck2007}:
\begin{equation}
\label{ham}
\begin{split}
&\hat{H}=\sum_{\ell} \sum_{m_{J}=-J}^{J} \Bigg[\,(\Delta_{\ell}+g_{J}(m_{J}+J)\Omega_{B})\,|\ell,m_{J}\rangle\langle\ell,m_{J}|\\
&{+}\;\!\bigg(\sum_{\substack{\{\ell',m_{J'}\} \\ >\,\{\ell,m_{J}\}}}{\sum_{\sigma}}\frac{\Omega_{\ell,\ell'}}{2}CG^{\sigma,m_{J},m_{J'}}_{\ell,\ell'}\Big(\frac{|\sigma|}{\sqrt{2}}\,\text{sin}(\alpha_{\ell,\ell'})\,e^{-i(\sigma\upvarphi_{\ell,\ell'}+\,\pi/2)}\\
&{+}\;\!(1-|\sigma|\,)\,\text{cos}(\alpha_{\ell,\ell'})\Big)\,|\ell',m_{J'}\rangle\langle\ell,m_{J}|\,\bigg) + h.c. \,\Bigg].\\
\end{split}
\end{equation}
The first term in the brackets represents the energy shift of the Zeeman state $|\ell,m_{J}\rangle$ due to the laser detuning and magnetic field, where $g_{J}$ is the Land\'{e} $g$-factor.
The second term and its hermitian conjugate contain the off-diagonal couplings.
We denote the photon polarization basis states by $|\sigma\rangle$, with $\sigma\in\{-1,0,1\}$.
The laser connecting $|\ell\rangle$ to $|\ell'\rangle$ propagates at an angle $\upvarphi_{\ell,\ell'}$ relative to the $x$-axis and is linearly polarized at an angle $\alpha_{\ell,\ell'}$ from the $z$-axis.
These two angles control the projection of the laser photon's angular momentum onto the quantization axis and the relative phase of the Rabi frequencies, $\Omega_{\ell,\ell'}$.
For a coherent process, these phases are well-defined and we take the $\Omega_{\ell,\ell'}$ to be real valued.
The Clebsch-Gordan coefficients, $CG^{\sigma,m_{J},m_{J'}}_{\ell,\ell'}$, set the relative strength of the drive between different magnetic sublevels of $|\ell\rangle$ and $|\ell'\rangle$.

We incorporate the non-Hermitian component of the 3-photon process using the density matrix formalism.
The density matrix for our system is
\begin{equation}
\label{rho}
\hat{\rho} = \sum_{\{\ell,m_{J}\}}\sum_{\{\ell',m_{J'}\}}\rho_{\scalebox{0.9}{$\scriptstyle{\{\ell,m_{J}\},\,\{\ell',m_{J'}\}}$}}|\ell,m_{J}\rangle\langle\ell',m_{J'}|.
\end{equation}
Each state's population, $\rho_{\scalebox{0.9}{$\scriptstyle{\{\ell,m_{J}\},\,\{\ell,m_{J}\}}$}}$, decreases at a rate given by the sum of the natural decay rates, $\Gamma_{\ell} = \sum_{\ell'}\,\Gamma_{\ell',\ell}$, connecting it to lower states $|\ell'\rangle$.
The coherence, $\rho_{\scalebox{0.9}{$\scriptstyle{\{\ell,m_{J}\},\,\{\ell',m_{J'}\}}$}}$, between states with distinct term symbols $\ell$ and $\ell'$ decays at a rate $\Gamma_{\ell,\ell'}/2$.
Coherences between the magnetic sublevels, $|\ell,m_{1}\rangle$ and $|\ell,m_{2}\rangle$ of a given $|\ell\rangle$ increase at a rate given by the decay of upper states into $|\ell\rangle$ and the Clebsch-Gordan coefficients governing the branching of those decays into $|\ell,m_{1}\rangle$ and $|\ell,m_{2}\rangle$.
In order to account for all of these processes, we define 
\begin{equation}
\label{helper}
\hat{\xi}^{}_{\sigma,\ell,\ell'} = \sum_{m,\,m'}CG^{\sigma,m_{J},m_{J'}}_{\ell,\ell'}|\ell,m_{J}\rangle\langle\ell',m_{J'}|,
\end{equation}
which allows us to construct the Liouville operator between $|\ell\rangle$ and $|\ell'\rangle$,
\begin{equation}
\label{liouville}
\begin{split}
&\hat{\mathcal{L}}_{\,\ell,\ell'} = \frac{\Gamma_{\ell,\ell'}}{2}\sum_{\sigma}\Big[\big(\hat{\xi}^{}_{\sigma,\ell,\ell'}\hat{\rho}\;\!\hat{\xi}^{\dagger}_{\sigma,\ell,\ell'}-\hat{\xi}^{\dagger}_{\sigma,\ell,\ell'}\hat{\xi}^{}_{\sigma,\ell,\ell'}\hat{\rho}\big)+h.c.\Big].
\end{split}
\end{equation}

\begin{figure}
\includegraphics[width=\linewidth]{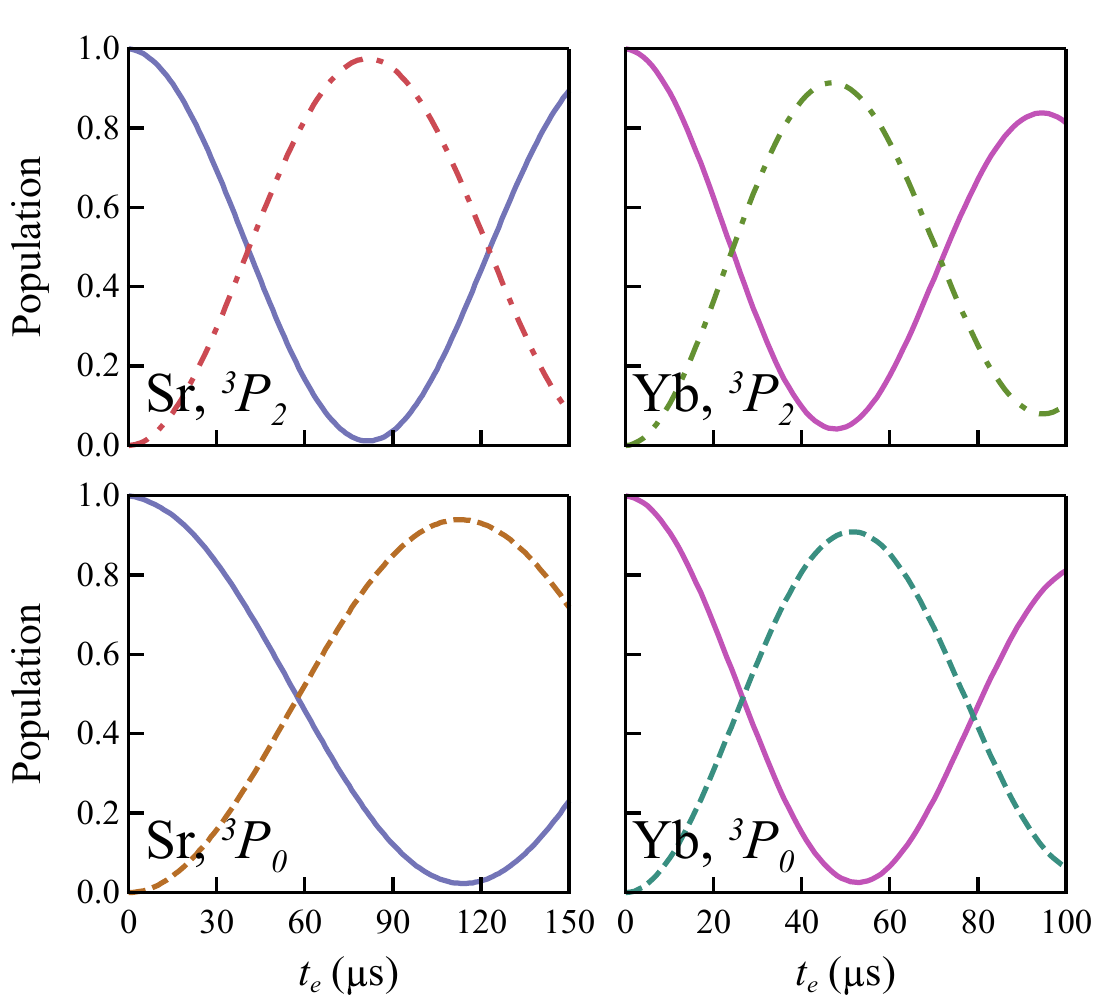}
\caption{\label{rabi}(Color Online) Results of the numerical integration of the Optical Bloch Equations (OBEs) for bosonic AE atoms given in equation~(\ref{OBE}). Solid curves show ground state populations while dashed and dash-dotted curves show target state populations ($|^{3}P_{0},0\rangle$ and $|^{3}P_{2},-2\rangle$, respectively). The populations in other states are negligible and not shown. Each laser turns on instantaneously at time zero and is instantaneously extinguished at time $t_{e}$. All the parameters for these simulations can be found in Table~\ref{simpars}.}
\end{figure}

By combining equations~(\ref{ham}) and~(\ref{liouville}), we arrive at the OBEs for the system~\cite{Steck2007}
\begin{equation}
\label{OBE}
\frac{d}{dt}\,\hat{\rho} = -i\,\big[\hat{H},\hat{\rho}\big]+\sum_{\ell,\ell'}\hat{\mathcal{L}}_{\ell,\ell'}.
\end{equation}
Once we insert the propagation angles, $\upvarphi_{\ell,\ell'}$, and the polarization angles, $\alpha_{\ell,\ell'}$, for the target state (either $|\term{3}{P}{2},\pm2\rangle$ or $|\term{3}{P}{0},0\rangle$, see Table~\ref{simpars}), the OBEs contain seven free parameters.
These are the detunings ($\Delta_{\term{3}{P}{1}}$, $\Delta_{\term{3}{S}{1}}$, and $\Delta_{\term{3}{P}{0\,(2)}}$), the magnetic field strength, $\Omega_{B}$, and the three Rabi frequencies ($\Omega_{\term{1}{S}{0},\term{3}{P}{1}}$, $\Omega_{\term{3}{P}{1},\term{3}{S}{1}}$, and $\Omega_{\term{3}{P}{0\,(2)},\term{3}{S}{1}}$).
Our objective is to find experimentally reasonable values for the free parameters that produce Rabi dynamics between $|\term{1}{S}{0},0\rangle$ and the target state.

\begin{table}
\caption{\label{simpars}The input parameters and results for the simulations in Figure~\ref{rabi}. The peak population in the target state, $\rho_{\pi}$, is reached after an evolution time, $t_{\pi}$. By varying $\Delta_{^{3}P_{0\,(2)}}$ in the simulations, we can extract the full width at half maximum of the 3-photon resonance, $\gamma_{3\text{-}photon}$ (see Figure~\ref{lw}). We quote detunings in units of the natural decay rate of their associated level.}
\scriptsize
\begin{ruledtabular}
\begin{tabular}{l|cccc}
& Sr, $|^{3}P_{2},\pm2\rangle$ & Sr, $|^{3}P_{0},0\rangle$ & Yb, $|^{3}P_{2},\pm2\rangle$ & Yb, $|^{3}P_{0},0\rangle$ \\ [0.5 ex]
\hline
$\Delta_{\term{3}{P}{1}}$ & $-100\times\Gamma_{\term{3}{P}{1}}$ & $-470\times\Gamma_{\term{3}{P}{1}}$ & $-230\times\Gamma_{\term{3}{P}{1}}$ & $-210\times\Gamma_{\term{3}{P}{1}}$\T \\ [1.25 ex]
$\Delta_{\term{3}{S}{1}}$ & $-21.6\times\Gamma_{\term{3}{S}{1}}$ & $-86.4\times\Gamma_{\term{3}{S}{1}}$ & $-38.9\times\Gamma_{\term{3}{S}{1}}$ & $-47.2\times\Gamma_{\term{3}{S}{1}}$ \\ [1.25 ex]
$\Delta_{\term{3}{P}{0\,(2)}}$ & 0 & 0 & 0 & 0 \\ [1.25 ex]
$\Omega_{\term{1}{S}{0},\term{3}{P}{1}}$ & $2\pi\times0.25\MHz$ & $2\pi\times0.25\MHz$ & $2\pi\times1.5\MHz$ & $2\pi\times1.5\MHz$ \\ [1.25 ex]
$\Omega_{\term{3}{P}{1},\term{3}{S}{1}}$ & $2\pi\times100\MHz$ & $2\pi\times90\MHz$ & $2\pi\times150\MHz$ & $2\pi\times150\MHz$ \\ [1.25 ex]
$\Omega_{\term{3}{P}{0\,(2)},\term{3}{S}{1}}$ & $2\pi\times4.0\MHz$ & $2\pi\times3.0\MHz$ & $2\pi\times6.0\MHz$ & $2\pi\times4.1\MHz$ \\ [1.25 ex]
$\Omega_{B}$ & $2\pi\times10\MHz$ & $2\pi\times10\MHz$ & $2\pi\times10\MHz$ & $2\pi\times10\MHz$ \\ [0.75 ex]
$\alpha_{\term{1}{S}{0},\term{3}{P}{1}}$ & $90\degree$ & $90\degree$ & $90\degree$ & $90\degree$ \\ [0.75 ex]
$\alpha_{\term{3}{P}{1},\term{3}{S}{1}}$ & $0\degree$ & $0\degree$ & $0\degree$ & $0\degree$ \\ [0.75 ex]
$\alpha_{\term{3}{P}{0\,(2)},\term{3}{S}{1}}$ & $90\degree$ & $90\degree$ & $90\degree$ & $90\degree$ \\ [0.75 ex]
$\upvarphi_{\term{1}{S}{0},\term{3}{P}{1}}$ & $-60.9\degree$ & $-59.6\degree$ & $-53.8\degree$ & $-51.5\degree$ \\ [0.75 ex]
$\upvarphi_{\term{3}{P}{1},\term{3}{S}{1}}$ & $60.8\degree$ & $59.5\degree$ & $80.7\degree$ & $73.3\degree$ \\ [1.0 ex]
\hline
$\rho_{\pi}$ & $97.5\%$ & $94.0\%$ & $91.5\%$ & $91.0\%$\T \\ [0.5 ex]
$t_{\pi}$ & $81.5\us$ & $112.3\us$ & $46.9\us$ & $51.4\us$ \\ [0.5 ex]
$\gamma_{3\text{-}photon}$ & $10.0\kHz$ & $7.2\kHz$ & $17.6\kHz$ & $15.5\kHz$ \\
\end{tabular}
\end{ruledtabular}
\end{table}

\begin{figure}
\includegraphics[width=\linewidth]{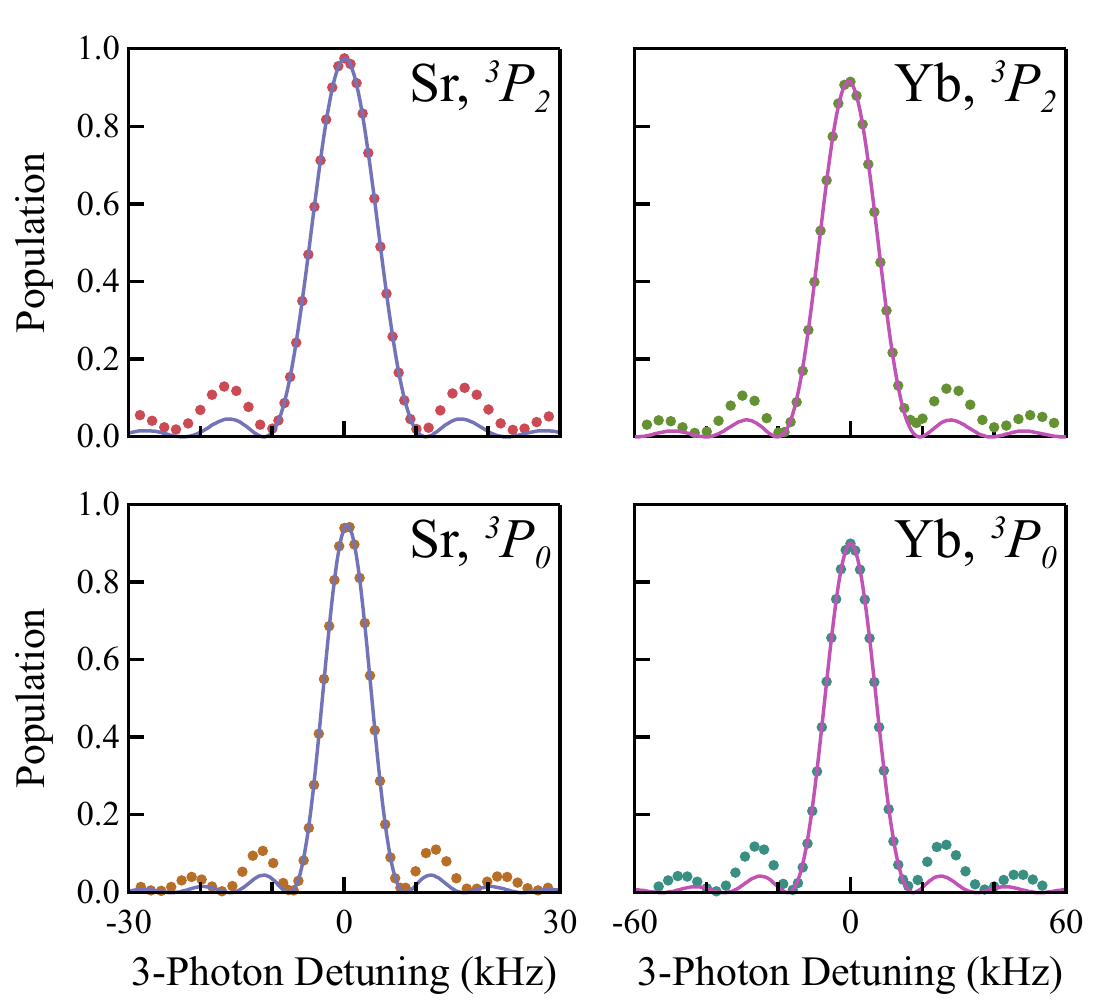}
\caption{\label{lw}(Color Online) The excitation fraction of the target state, $|^{3}P_{2},\pm2\rangle$ ($|^{3}P_{0},0\rangle$), at $t_{\pi}$ as a function of the 3-photon detuning is shown in the top (bottom) row. The left column shows the results for Sr and the right column shows the results for Yb. Except for the 3-photon detuning, which is varied, the parameters for the simulations are identical to those used in Figure~\ref{rabi} and reported in Table~\ref{simpars}. The $\text{sinc}^{2}$ fits (solid lines) yield full widths at half maximum for the 3-photon transitions to $^{3}P_{2}\,(^{3}P_{0})$ of $10.0\,(7.2)\kHz$ for Sr and of $17.6\,(15.5)\kHz$ for Yb. The amplitude mismatch between the wings of the fit and the simulation is likely caused by the convolution of power broadening and the transform limit.}
\end{figure}

We numerically integrate the OBEs and vary the input parameters to optimize the amplitude of Rabi oscillations.
The optical fields all turn on instantaneously at time zero and uniformly illuminate the system for an evolution time, $t_{e}$, at which point they are all instantaneously extinguished.
For both Sr and Yb, we find values of the detunings and couplings that yield $\gtrsim90\%$ peak population transfer.
Figure~\ref{rabi} shows the evolution of the relevant diagonal elements of the density matrix for these parameter sets.
The transfer efficiency is slightly higher for strontium than ytterbium, which is likely due to the reduced linewidth of the $\term{1}{S}{0}\rightarrow\term{3}{P}{1}$ line.
The coherence of the dynamics allows a degenerate gas to be transferred to a metastable state with minimal heating.
We plot the peak excitation fraction of the target state as a function of the 3-photon detuning in Figure~\ref{lw}.
The width of these curves is a numerical estimate of the 3-photon linewidth, which we extract from a $\text{sinc}^{2}$ fit to the numerical results.
The fit captures the behavior of the central peak, but deviates in the wings because the actual lineshape is a convolution of multiple broadening effects.
We find that the full width at half maximum is $\simeq10\kHz$ for Sr and $\simeq20\kHz$ for Yb.
We have not simulated the 3-photon dynamics for AE fermions due to the $10\times$ ($6\times$, $2\times$) larger Hilbert space for $^{87}\text{Sr}$ ($^{173}\text{Yb}$, $^{171}\text{Yb}$).
However, the results for bosons suggest that high efficiency transfer of a spin-polarized Fermi degenerate gas to a metastable state is possible.

\section{\label{exp}Experimental Considerations}

In Section~\ref{thry}, we demonstrated that a coherent, 3-photon excitation scheme can theoretically transfer a large population fraction from $|\term{1}{S}{0},0\rangle$ to $|\term{3}{P}{0},0\rangle$ or $|\term{3}{P}{2},-2\rangle$.
However, there are several technical details that must be considered in order to realize the 3-photon process experimentally.
For example, the incident laser beams must be carefully aligned to minimize the net momentum transfer to the degenerate gas during excitation.
The differential AC Stark shift between the four Zeeman levels involved in the process must also be controlled to avoid inhomogeneous broadening of the 3-photon transition.
The elastic and inelastic interactions of the degenerate gas could also broaden the 3-photon transition or affect the utility of the resulting metastable sample.

For the 3-photon process to be successful, it must not excite center of mass oscillations or excessively heat the sample.
During population transfer, an atom emits a photon into the laser beam addressing $\term{3}{P}{0\,(2)}\rightarrow\term{3}{S}{1}$ and absorbs one photon each from the other two lasers.
Because three photons are involved in the excitation, we can find angles of incidence, $\upvarphi_{\ell,\ell'}$, that eliminate net momentum transfer to the degenerate gas even though each photon carries distinct momentum.
For the laser configuration in Figure~\ref{scheme}, the $\upvarphi_{\ell,\ell'}$ are given by
\begin{equation}
\label{angs}
\begin{split}
& k_{\term{1}{S}{0},\term{3}{P}{1}}\text{cos}(\upvarphi_{\term{1}{S}{0},\term{3}{P}{1}})+k_{\term{3}{P}{1},\term{3}{S}{1}}\text{cos}(\upvarphi_{\term{3}{P}{1},\term{3}{S}{1}})=k_{\term{3}{S}{1},\term{3}{P}{0\,(2)}} \\
& k_{\term{1}{S}{0},\term{3}{P}{1}}\text{sin}(\upvarphi_{\term{1}{S}{0},\term{3}{P}{1}})+k_{\term{3}{P}{1},\term{3}{S}{1}}\text{sin}(\upvarphi_{\term{3}{P}{1},\term{3}{S}{1}})=0,
\end{split}
\end{equation}
where $k_{\ell,\ell'}$ is the wavenumber for the $|\ell\rangle\rightarrow|\ell'\rangle$ transition.
The first equation in~(\ref{angs}) represents the momentum transfer along $\hat{x}$ and the second is the transfer along $\hat{y}$ (see Figure~\ref{scheme}).
The angles for excitation to $\term{3}{P}{0\,(2)}$ for both Sr and Yb are given in Table~\ref{simpars}.
Misalignment of any of the lasers will lead to heating during the 3-photon transfer.
We can estimate the effect of misalignment by comparing the recoil energy of the net momentum after the 3-photon process ($E_{r}$) to the level spacing of the harmonic trapping potential ($E_{trap}$).
If we assume a trap frequency of $50\Hz$ and a $\pm1\degree$ error in either or both of $\upvarphi_{\term{1}{S}{0},\term{3}{P}{1}}$ and $\upvarphi_{\term{3}{P}{1},\term{3}{S}{1}}$, then $E_{r}/E_{trap}<0.09\,(0.07)$ for both target states in Sr (Yb).
Each of the lasers could also be misaligned out of the $xy-$plane. For the worst combination of $\pm1\degree$ vertical alignment errors and the same $50\Hz$ trap frequency, $E_{r}/E_{trap}<0.27\,(0.17)$ for either target state in Sr (Yb). Both types of misalignment result in $E_{r}/E_{trap}$ substantially less than unity, so the recoil heating should be insignificant.

\begin{table}[b]
\caption{\label{lambda}The wavelengths and linewidths for the transitions involved in the 3-photon process~\cite{NIST_ASD, Porsev1999}. Except for the $^{1}S_{0}\rightarrow\,^{3}P_{1}$ transition in Yb, diode lasers can easily generate the requisite wavelengths.}
\begin{ruledtabular}
\begin{tabular}{c|cccc}
\multirow{2}{*}{$\ell,\ell'$} & \multicolumn{2}{c}{Sr} & \multicolumn{2}{c}{Yb} \\
& $\lambda_{\ell,\ell'}$ & $\gamma_{\ell,\ell'}$ & $\lambda_{\ell,\ell'}$ & $\gamma_{\ell,\ell'}$ \\ [0.5 ex]
\hline
$\term{1}{S}{0},\term{3}{P}{1}$ & $689\nm$ & $7.5\kHz$ & $556\nm$ & $180\kHz$ \T \\ [0.25 ex]
$\term{3}{P}{0},\term{3}{S}{1}$ & $679\nm$ & $1.4\MHz$ & $649\nm$ & $1.5\MHz$ \\ [0.25 ex]
$\term{3}{P}{1},\term{3}{S}{1}$ & $688\nm$ & $4.3\MHz$ & $680\nm$ & $4.3\MHz$ \\ [0.25 ex]
$\term{3}{P}{2},\term{3}{S}{1}$ & $707\nm$ & $6.7\MHz$ & $770\nm$ & $6.0\MHz$ \\ [0.25 ex]
\end{tabular}
\end{ruledtabular}
\end{table}

The Rabi frequencies $\{\Omega_{\term{1}{S}{0},\term{3}{P}{1}}$, $\Omega_{\term{3}{P}{1},\term{3}{S}{1}}$, $\Omega_{\term{3}{P}{0},\term{3}{S}{1}}$, $\Omega_{\term{3}{P}{2},\term{3}{S}{1}}\}$ from Figure~\ref{rabi} correspond to saturation parameters, $I/I_{sat}$, on the order of $\{1000, 1000, 10, 1\}$ (For Yb, the saturation for $\Omega_{\term{1}{S}{0},\term{3}{P}{1}}$ is on the order of $100$).
For the $\term{3}{P}{1}\rightarrow\term{3}{S}{1}$ transition, a laser beam can achieve the necessary intensity with approximately $10\mW$ of power and a $1/e^{2}$ radius $\simeq400\um$.
The other transitions only require a beam with $\lesssim1\mW$ of power and a waist $\simeq3\mm$ to reach the appropriate saturation.
Diode lasers can easily produce these powers and the requisite waists are much larger than the typical dimensions of a degenerate gas, which will suppress dephasing due to the gaussian intensity profile of the laser beams.

The three excitation lasers must be phase stabilized to better than the 3-photon linewidth (see Figure~\ref{lw} and Table~\ref{simpars}) in order to produce coherent dynamics. 
The lasers for Sr have very similar wavelengths (see Table~\ref{lambda}), and interrogation of the $\term{1}{S}{0}\rightarrow\term{3}{P}{1}$ transition in Sr typically requires a high finesse optical cavity to decrease that laser's linewidth.
This makes a cavity transfer lock an appealing strategy, and cavity mediated stability transfer at the $\lesssim10\kHz$ level has been demonstrated in the context of Sr Rydberg excitation~\cite{Bridge2016}.
The excitation wavelengths span a much larger range for Yb, which increases the technical difficulty of a cavity transfer lock.
The lasers could instead be locked using a combination of cascade and lambda type EIT~\cite{Bell2007,Abel2009} or by stabilizing each laser with an optical frequency comb~\cite{Inaba2013,Scharnhorst2015}.

Ideally, the 3-photon excitation would occur in an optical trap with no differential AC stark shift between any of the coupled levels.
However, due to the dipole-allowed transitions between $|\term{3}{S}{1}\rangle$ and the states in the $\term{3}{P}{J}$ manifold, we should expect that no practical wavelength satisfies this condition.
Intuition from 2-photon Raman processes suggests, and simulations of our system verify, that trap induced shifts to the intermediate detunings ($\Delta_{\term{3}{P}{1}}$ and $\Delta_{\term{3}{S}{1}}$) contribute only weakly to inhomogeneous broadening of the 3-photon transition.
This allows population transfer to occur in a trap operating at a magic wavelength that eliminates the differential AC Stark shift between the initial and final states.
The magic wavelengths for $|\term{1}{S}{0}\rangle$ and $|\term{3}{P}{0}\rangle$ are well known in Sr and Yb because of their application to optical clocks~\cite{Nicholson2015,Hinkley2013,Bloom2014}.
To search for magic wavelengths for $|\term{3}{P}{2},\pm2\rangle$, we calculate the polarizability (scalar and tensor) for each state involved in the multi-photon transition following the procedure in~\cite{Boyd2007,Steck2007} with lines from~\cite{NIST_ASD} and linewidths from~\cite{NIST_ASD,Werij1992,Mickelson2005,Miller1992,Andra1975,Parkinson1976,Ido2003}, for Sr, and~\cite{NIST_ASD,Porsev1999,Takasu2004,Dzuba2010}, for Yb.
We expect the calculated polarizabilities to predict magic wavelengths with better than $\pm10\nm$ accuracy (the level at which it reproduces known magic wavelengths in Sr) except for the Yb $\term{3}{P}{0}$ state, for which few matrix elements are reported in the literature.

\begin{figure}[!]
\includegraphics[width=\linewidth]{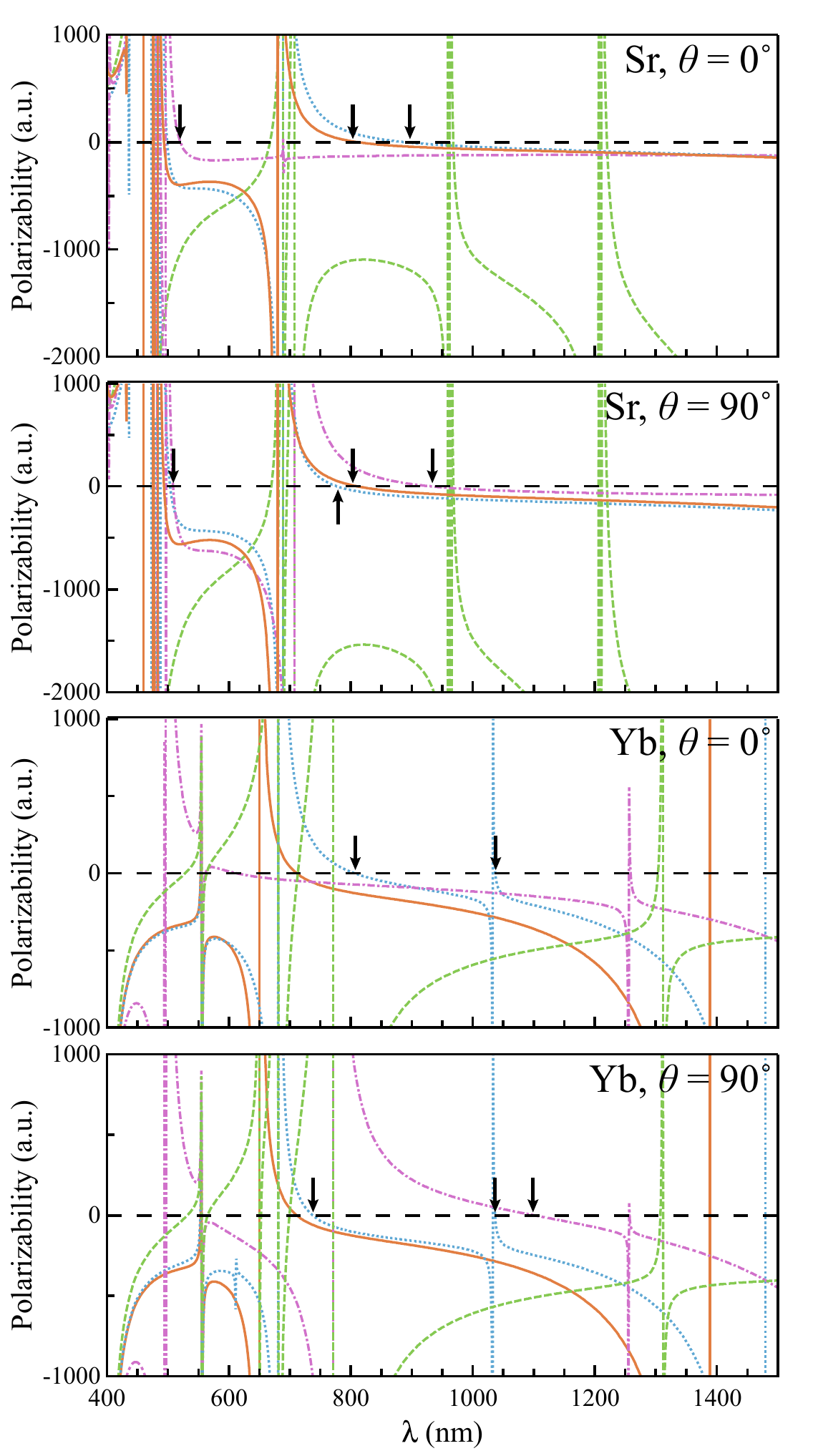}
\caption{\label{pol}(Color Online) The AC polarizabilities of $|^{3}P_{0},0\rangle$ (solid orange), $|^{3}P_{1},\pm1\rangle$ (dotted blue), $|^{3}P_{2},\pm2\rangle$ (dash-dotted magenta), and $|^{3}S_{1},\pm1\rangle$ (dashed green) relative to $|^{1}S_{0},0\rangle$ in atomic units. The top two panels show the polarizability of Sr for ODT light polarized parallel and perpendicular to the quantization axis, respectively. The bottom two panels are the corresponding plots for Yb. Every zero crossing indicates a magic wavelength. We indicate the magic wavelengths of specific interest here using black arrows. For $|^{3}P_{2},\pm2\rangle$ and $|^{3}P_{1},\pm1\rangle$, the laser polarization can tune the indicated magic wavelengths over a wide range (see Table~\ref{magic}). Note that the position of the $|^{3}P_{0},0\rangle$ and $|^{3}P_{2},\pm2\rangle$ magic wavelengths flips with respect to the $|^{3}P_{1},\pm1\rangle$ magic wavelength in the near-IR as the polarization angle changes from $0\degree$ to $90\degree$. This indicates the presence of a double magic wavelength at a particular polarization angle that equalizes the AC Stark shifts of three energy levels.}
\end{figure}

Figure~\ref{pol} contains the results of our calculations for trapping lasers polarized parallel and perpendicular to the magnetic field axis.
The plots show the differential polarizability between $\{|\term{3}{P}{0},0\rangle,\,|\term{3}{P}{1},\pm1\rangle,\,|\term{3}{P}{2},\pm2\rangle,\,|\term{3}{S}{1},\pm1\rangle\}$ and the ground state, $|\term{1}{S}{0},0\rangle$.
In the two upper panels of Figure~\ref{pol}, we see two magic wavelengths for $|\term{1}{S}{0},0\rangle$ and $|\term{3}{P}{2},\pm2\rangle$ near $520\nm$ and $950\nm$.
The lower panels in Figure~\ref{pol} indicate that in Yb there is a magic wavelength for these two states near $1100\nm$.
All of these magic wavelengths can be tuned over a wide range ($\gtrsim100\nm$ for the near-IR wavelengths) by varying the dipole trap polarization angle, $\theta$, between $0\degree$ and $90\degree$.
In particular, the Yb $|\term{3}{P}{2},\pm2\rangle$ magic wavelength moves to $1064\nm$ when $\theta\approx66\degree$ and the green magic wavelength in Sr is tunable over the range $\{508,520\}\nm$.
We can also see in Figure~\ref{pol} that the near-IR magic wavelengths for $|\term{3}{P}{1},\pm1\rangle$ and $|\term{3}{P}{2},\pm2\rangle$ have opposite angular dependence in both Sr and Yb.
This observation suggests the existence of a doubly-magic wavelength, $\lambda_{\scalebox{0.85}{$\scriptstyle{2\times}$}m}$, that eliminates differential light shifts between $|\term{3}{P}{2},\pm2\rangle$, $|\term{3}{P}{1},\pm1\rangle$ and $|\term{1}{S}{0},0\rangle$ when the dipole trap is polarized at a magic angle, $\theta_{\scalebox{0.85}{$\scriptstyle{2\times}$}m}$.
By varying $\theta$, we are able to identify one doubly-magic wavelength in Sr and two in Yb (see Table~\ref{magic}), which would allow further reduction of the trap-induced inhomogeneous broadening.
We note that the optical clock transition magic wavelengths can also be made doubly-magic for both elements by tuning the polarizability of $|\term{3}{P}{1},\pm1\rangle$.

The remaining AC Stark shift of $|\term{3}{S}{1}\rangle$ or $|\term{3}{P}{1}\rangle$ with respect to the ground state inhomogeneously broadens the 3-photon transition.
This inhomogeneous broadening arises due to the different harmonic confinement of $|\term{3}{S}{1}\rangle$ and $|\term{3}{P}{1}\rangle$ compared to the ground state. 
We estimate the scale of the broadening by taking the difference between the ground state chemical potential of the degenerate gas, $\mu_{\term{1}{S}{0}}$, and the chemical potential it would have in $|\term{3}{S}{1}\rangle$, $\mu_{\term{3}{S}{1}}$, or $|\term{3}{P}{1}\rangle$, $\mu_{\term{3}{P}{1}}$.
Typical degenerate gases have chemical potentials on the order of $1\kHz$.
The excited state chemical potential is related to the ground state chemical potential by the ratio of the polarizabilities of the two states, assuming that the $s$-wave scattering lengths are equal.
For the magic wavelengths under consideration (see Table~\ref{magic}), the ratio of the $|\term{3}{S}{1},\pm1\rangle$ and $|\term{1}{S}{0},0\rangle$ polarizabilities for Sr (Yb) ranges from $-0.1$ to $-10$ ($-2$ to $-40$).
This means that the inhomogeneous broadening, $\mu_{\term{1}{S}{0}}-\mu_{\term{3}{S}{1}}$, should be $\lesssim10\kHz$ for Sr and $\lesssim40\kHz$ for Yb.
Because the inhomogeneous broadening due to $|\term{3}{S}{1},\pm1\rangle$ is substantially smaller than the 2-photon detuning, its effect on the Rabi dynamics should be negligible. 
Similarly, our reasoning implies that the inhomogeneous broadening from $|\term{3}{P}{1},\pm1\rangle$ will be $\lesssim5\kHz$, which is substantially smaller than the 1-photon detunings we consider.

\begin{table}
\caption{\label{magic} Magic wavelengths and magic wavelength ranges for the target states in the 3-photon process. We also list the double magic wavelengths and associated angles that equalize the AC Stark shifts of $|^{3}S_{0},0\rangle$, $|^{3}P_{1},\pm1\rangle$, and $|^{3}P_{2},\pm2\rangle$ or $|^{3}P_{0},0\rangle$. For $|^{3}P_{0},0\rangle$, magic wavelengths for fermionic isotopes are taken from~\cite{Hinkley2013,Nicholson2015} and rounded to the nearest nm.}
\begin{ruledtabular}
\begin{tabular}{c|cccc}
& Sr, $|^{3}P_{2},\pm2\rangle$ & Sr, $|^{3}P_{0},0\rangle$ & Yb, $|^{3}P_{2},\pm2\rangle$ & Yb, $|^{3}P_{0},0\rangle$ \\ [0.25 ex]
\hline
$\lambda_{magic}$ & \makecell{$720\;\!{-}\;\!935\nm$\T \\ $508\;\!{-}\;\!520\nm$} & $813\nm$ & $780\;\!{-}\;\!1100\nm$ & $759\nm$ \\ 
\hline
$\lambda_{\scalebox{0.85}{$\scriptstyle{2\times}$}m}$ & $823\nm$ & $813\nm$ & \makecell{$802\nm$\T \\ $1036\nm$} & $759\nm$ \\ 
$\theta_{\scalebox{0.85}{$\scriptstyle{2\times}$}m}$ & $44\degree$ & $55\degree$ & \makecell{$15\degree$\T \\ $58\degree$} & $55\degree$ \\
\end{tabular}
\end{ruledtabular}
\end{table}

The $s$-wave scattering length of atoms in the metastable degenerate gas will generally differ from the scattering length of atoms in the ground state.
This difference in interaction strength will cause both a shift in the resonance frequency and inhomogeneous broadening of the 3-photon transition.
There have been few measurements or calculations of the scattering lengths of metastable AE atoms~\cite{Derevianko2003,Zhang2014,Cappellini2014a,Scazza2014}, so an accurate estimate of this inhomogeneous broadening is not possible.
However, it is reasonable to assume that the interaction broadening will be on the order of the ground state chemical potential, $\mu_{\term{1}{S}{0}}$, and thus only weakly perturb the 3-photon dynamics since the intermediate detunings are large.
The inelastic collision rate for metastable AE atoms is on the order of $10^{-10}-10^{-11}~\text{cm}^{3}/\text{s}$ depending on the state and species~\cite{Traverso2009, Lisdat2009, Yamaguchi2008,Uetake2012,Halder2013}, while AE degenerate gases usually have densities $\gtrsim10^{13}~\text{cm}^{-3}$~\cite{Takasu2003,Stellmer2014a}.
We would thus expect the lifetime of the metastable degenerate gas to be limited to $\simeq10\ms$, severely restricting the experimental timescale.
However, several AE atom isotopes ($^{40}\text{Ca}$, $^{86}\text{Sr}$, $^{168}\text{Yb}$) have sufficiently strong interactions to allow cooling to degeneracy at relatively low densities ($\simeq10^{12}~\text{cm}^{-3}$)~\cite{Stellmer2010,Halder2012a,Sugawa2011}.
These isotopes could also sympathetically cool other isotopes to degeneracy at low density.
Alternatively, recent advances in optical trapping techniques might allow dynamic decompression of the ground state degenerate gas at fixed trap depth~\cite{Roy2016,Bell2016}.
By using either a strongly interacting isotope or a dynamically decompressed trap to generate a low density sample, the lifetime of a metastable degenerate gas could be extended to $\simeq100\ms$.
This timescale is sufficient for a wide variety of experiments, and for $|\term{3}{P}{2}\rangle$ it could be extended even further using a magnetic Feshbach resonance~\cite{Derevianko2003}.
The Bose-Einstein or Fermi-Dirac statistics of the degenerate gas will suppress inelastic collisions due to changes in the 2-particle correlation function (as has been observed for three-body loss processes in, \textit{e.g.},~\cite{Burt1997,Ottenstein2008}), potentially allowing longer sample lifetimes.

\section{\label{con}Conclusions}

We have proposed and studied a coherent 3-photon process for creating quantum degenerate metastable samples of AE atoms.
Numerical simulations of the 3-photon Rabi dynamics show that $\simeq90~\%$ population transfer to $\term{3}{P}{0\,(2)}$ can be achieved in Sr and Yb.
Similar transfer efficiency should be attainable in Ca as well.
The smaller mass of Ca reduces the linewidth of the $^{1}S_{0}\rightarrow\,^{3}P_{1}$ transition (to $\simeq370\Hz$), which will increase the technical challenge of phase locking the necessary lasers.
The Rabi dynamics are fast compared to reasonable trap oscillation frequencies, but slow compared to typical experimental timing resolution.
We considered several experimental obstacles to implementation of the transfer scheme. The excitation lasers require moderate laser power ($\lesssim10\mW$) with reasonable beam waists ($\gtrsim100\um$) and can be arranged to cancel the momentum kick during population transfer.
An optical dipole trap could operate near a doubly magic wavelength to cancel the differential polarizability between three of the four states involved in the excitation.
Even in the worst case, the inhomogeneous broadening due to the remaining state, $\term{3}{S}{1}$, is insignificant.
The ratio of the detuning from, and linewidth of, the remaining state ($\term{3}{S}{1}$) to the worst case inhomogeneous broadening induced by the trap is large enough to render the broadening insignificant.
A similar argument applies to interaction induced broadening effects.
By using an isotope with a large ground state $s$-wave scattering length or by dynamically changing the trapping potential to decompress the degenerate gas at fixed trap depth, the lifetime of the final metastable sample could be extended to $\simeq100\ms$.
The Bose or Fermi statistics of the metastable sample suppress inelastic collisions~\cite{Burt1997,Ottenstein2008} and will increase the lifetime further (potentially to several seconds for spin-polarized Fermi degenerate gases).
The $100\ms$ timescale is long enough to perform useful experiments in the thermodynamically 3-dimensional limit~\cite{Herold2012} or to adiabatically ramp on an optical lattice to create a Mott insulating state for quantum simulation experiments~\cite{Bhongale2013,Lahrz2014}.

The authors thank A. Gorshkov, Z. Smith, V. Vaidya, and S. Eckel for useful discussions.
This work was partially supported by ONR, and the NSF through the PFC at the JQI.

\bibliography{3PhotonDynamics}

\end{document}